\documentclass[prb,twocolumn,showpacs,floatfix]{revtex4}
\usepackage{graphicx}
\usepackage{dcolumn}
\usepackage{bm}

\begin{document}

\preprint{}

\title{Manipulating the spin texture in spin-orbit superlattice by terahertz radiation}

\author{D.V. Khomitsky}
 \email{khomitsky@phys.unn.ru}
 \affiliation{Department of Physics, University of Nizhny Novgorod,
              23 Gagarin Avenue, 603950 Nizhny Novgorod, Russian Federation}

\date{\today}

\begin{abstract}
The spin texture in a gate-controlled one-dimensional superlattice with Rashba spin-orbit coupling
is studied in the presence of external terahertz radiation causing the superlattice
miniband transitions. It is shown that the local distribution of the excited spin density
can be modified by varying the Fermi level of the electron gas and by changing the radiation intensity
and polarization, allowing the controlled coupling of spins and photons.

\end{abstract}

\pacs{72.25.Fe, 73.21.Cd, 78.67.Pt}

\maketitle

\section{Introduction}

The control of the spin degrees of freedom is one of the primary goals of rapidly
developing field of condensed matter physics known as spintronics.\cite{Awschalom,Zutic}
In addition to the methods involving the magnetic field which effectively governs the
spins, the application of non-magnetic spin systems can be considered by taking into
account the spin-orbit (SO) interaction. In most widely considered two-dimensional semiconductor
heterostructures the SO interaction is usually dominated by the Rashba coupling\cite{Rash60} coming
from the structure inversion asymmetry of confining potential and effective mass difference.
The value of Rashba coupling strength can be tuned by the external gate voltage\cite{Miller}
and it reaches the value of $2 \cdot 10^{-11}$ eVm in InAs-based structures.\cite{Grundler}
One of the important issues of spintronics is the interaction between spins and
photons which is promising for further applications in novel electronic and optical
devices. The studies of optical properties of SO semiconductor structures have formed
a fast growing field of studies during the last decade. Some of the research topics
included the photogalvanic,\cite{Golub,Ganichev03,Ganichev07,Yang,Zhou,Cho}
spin-galvanic\cite{Ganichev03R,Ganichev07} and spin-photovoltaic\cite{Fedorov} effects as well as optical
spin orientation\cite{Tarasenko} and pure spin current generation.\cite{Bhat,Zhao}
The effects of terahertz radiation onto spin-split states in semiconductors were also the subject of
investigation.\cite{Olesberg,Khurgin} Another important property of non-uniform spin distributions such
as spin coherence standing waves was discussed by Pershin\cite{Pershin} who found an increase
of the spin relaxation time in such structures, making them promising for spintronics applications.
One of possible ways to create a non-uniform spin distribution in a heterostructure is to apply
a metal-gated superlattice with tunable amplitude of electric potential to the two-dimensional
electron gas (2DEG) with spin-orbit coupling. The quantum states and spin polarization in this system
with standing spin waves were studied previously,\cite{jetpl} and the problem of scattering
on such structure has been considered.\cite{soscat}

In the present Brief Report we study the problem of the excited spin density creation in
a one-dimensional InAs-based superlattice with Rashba spin-orbit coupling by applying
an external terahertz radiation. The direction of photon propagation is chosen perpendicular
to 2DEG plane where both linear and circular polarizations are considered. It is shown
that the excited spin texture is sensitive to the position of the Fermi level of the 2DEG and
to the radiation intensity. The former can be tuned by the gate voltage, thus providing a new possible
way to couple local excited distribution of spins and photons in 2DEG with SO interaction.
The obtained results have a qualitative character and are not restricted to one specific type
of semiconductor heterostructure, superlattice period, amplitude of periodic potential,
Fermi level position, etc. Another important issue is the spin relaxation which tends to transform
the excited spin distribution back to the equilibrium. The terahertz scale of excitation frequencies
is at least of an order of magnitude larger than the spin relaxation rates in InAs semiconductor
heterostructures which can be estimated as $1/\tau_s$ where the spin lifetime $\tau_s$ reaches there
the values from $60$ ps\cite{Murdin} to $600$ ps\cite{Hall} and the spin relaxation time is further
increased in non-homogeneous spin textures.\cite{Pershin} Thus, one can expect that the effects discussed
in the current Brief Report can be experimentally observable.

This Brief Report is organized as follows. In Sec.II we briefly describe quantum states in SO
superlattice and in Sec.III we calculate the spatial distribution of the excited spin density in
a superlattice cell under terahertz radiation with different positions of Fermi level,
radiation intensity and polarization. The conclusions are given in Sec.IV.

\section{Quantum states in SO superlattice}

We start with the brief description of the quantum states of 2DEG with Rashba SO coupling
and one-dimensional (1D) periodic superlattice potential in the absence of external
electromagnetic field.\cite{jetpl} The Hamiltonian is the sum of the 2DEG kinetic energy
operator in a single size quantization band with effective mass $m$, the Rashba SO term
with strength $\alpha$ and the periodic electrostatic potential of the 1D superlattice:

\begin{equation}
\hat{H}=\frac{\hat{p}^2}{2m}+\alpha(\hat{\sigma}_x\hat{p}_y-\hat{\sigma}_y\hat{p}_x)
        +V(x),
\label{ham}
\end{equation}

where $\hbar =1$ and the periodic potential is chosen in the simplest form
$V(x)=V_0\cos 2\pi x /a$ where $a$ is the superlattice period
and the amplitude $V_0$ is controlled by the gate voltage.
The eigenstates of Hamiltonian (\ref{ham}) are two-component Bloch spinors with
eigenvalues labeled by the quasimomentum $k_x$ in a one-dimensional Brillouin zone
$-\pi/a \le k_x \le \pi/a$, the momentum component $k_y$, and the miniband index $m$:

\begin{equation}
\psi_{m\bf k}=\sum_{\lambda n} a^m_{\lambda n} ({\bf k})
\frac{e^{i{\bf k}_n{\bf r}}}{\sqrt{2}} \left(
\begin{array}{c}
1 \\
\lambda e^{i\theta_n}
\end{array}
\right),
\quad \lambda = \pm 1.
\label{wf}
\end{equation}

\begin{figure}[ht]
  \centering
  \includegraphics[width=80mm]{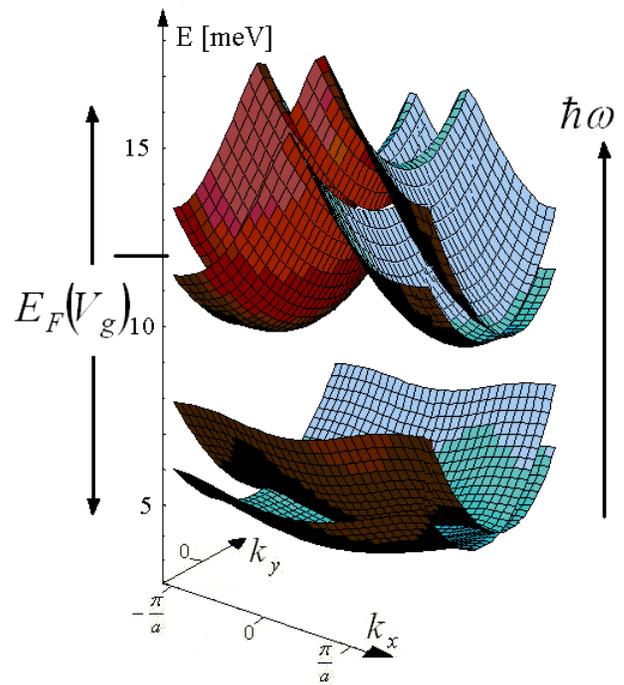}
  \caption{(color online) Energy spectrum of four lowest minibands in the InAs 1D superlattice with
           Rashba constant $\alpha=2 \cdot 10^{-11}$ eVm, the electron effective mass $m=0.036$ $m_0$,
           the superlattice period and amplitude $a=60$ nm and $V_0=10$ meV.
           The Fermi level position $E_F(V_g)$ and the photon energy
           $\hbar \omega=10$ meV corresponding to $\omega/2\pi=2.43$ THz are shown
           schematically.}
  \label{figband}
\end{figure}

Here ${\bf k}_n={\bf k}+n{\bf b}=\left(k_x+\frac{2\pi}{a}n, \quad k_y\right)$ and
$\theta_n={\rm arg}[k_y - ik_{nx}]$. The energy spectrum of Hamiltonian (\ref{ham})
consists of pairs of spin-split minibands determined by the SO strength $\alpha$
separated by the gaps of the order of $V_0$. An example of the energy spectrum is shown in Fig.\ref{figband}
for the four lowest minibands in the InAs 1D superlattice with Rashba constant $\alpha=2 \cdot 10^{-11}$ eVm,
the electron effective mass $m=0.036$ $m_0$, the superlattice period $a=60$ nm and the amplitude of
the periodic potential $V_0=10$ meV. It should be mentioned that the spectrum in Fig.\ref{figband}
is limited to the first Brillouin zone of the superlattice in the $k_x$ direction while
the cutoff in the $k_y$ direction is shown only to keep the limits along $k_x$ and $k_y$ comparable.
The Fermi level position $E_F(V_g)$ can be varied by tuning the gate voltage which controls
the concentration of the 2D electrons. This feature is shown in Fig.\ref{figband} schematically
by the arrows near $E_F(V_g)$ as well as the photon energy $\hbar \omega=10$ meV corresponding
to $\omega/2\pi=2.43$ THz.

\section{Spin texture manipulation}

The Rashba SO coupling as well as the periodic superlattice potential cannot produce
the net polarization of 2DEG. Moreover, the two-component eigenvectors of the Rashba
Hamiltonian describe the homogeneous local spin density $\sigma_i=\psi^{\dagger} {\hat \sigma}_i \psi$,
where $i=x,y,z$. In the presence of an additional superlattice potential, however,
the local spin density for a given state $(k_x,k_y)$ can be inhomogeneous,\cite{jetpl}
as in the spin coherence standing wave,\cite{Pershin}
which gives an idea to obtain a non-uniform spin density distribution under an external
radiation which involves in transitions the states with different $(k_x,k_y)$ with
a varying impact depending on the matrix elements. In this Section we subject the 2DEG with
Hamiltonian (\ref{ham}) to the electromagnetic radiation propagating along $z$ axis perpendicular
to the 2DEG plane with the electric field of the radiation
${\bf E}(t)={\bf e}E_{\omega}\exp^{-i\omega t} + \quad {\rm c.c.}$ with amplitude $E_{\omega}$,
frequency $\omega$ polarization ${\bf e}=(e_x,e_y)$.
When the electromagnetic radiation is applied, the excited spin density rate $S_i$
at a given point in a real space can be found in the following way:\cite{Bhat}

\begin{equation}
S_{i}=\frac{\pi e^2 E_{\omega}^2 }{\omega^2}
\int d^2k \sum_{jl}\xi_i^{jl}({\bf k}){\bar e}_je_l
\label{s}
\end{equation}

where

\begin{eqnarray}
\xi_i^{jl}({\bf k})=\sum_{c,m,m'}
\left(\psi_{m'}^{\dagger}\hat{\sigma_i}\psi_{m}\right)
{\bar v}^j_{m'c}v^l_{mc}
\\
\times\left[\delta(\omega_{mc}({\bf k})-\omega)+
\delta(\omega_{m'c}({\bf k})-\omega)\right].
\label{xi}
\end{eqnarray}

Since the structure is completely homogeneous in the $y$ direction, the $S_y$ component vanishes.
The other components $S_x$ and $S_z$ of the excited non-equilibrium spin density can be nonzero
at a given point in a superlattice even for the linear $x$-polarized radiation. The coexistence of
the axial vector components $(S_x,S_z)$ in the left side of Eq.(\ref{s}) together with the polar vector
component $e_x$ in the right side is in agreement with the principles of magnetic crystal class
analysis.\cite{LL} There is a mirror plane of reflection $\sigma_y$ in our system which changes
$y$ to $-y$ and thus changes the sign of the magnetic moment, leaving the $e_x$ component of
the polarization unchanged. The element of the magnetic crystal class, however, is applied only as
a combination $\sigma_{y}R$ where $R$ is the time reversal operator\cite{LL} which again changes
the direction of the magnetic moment but does not change the polarization component $e_x$. As a result,
the combination $\sigma_{y}R$ leaves both spin projections $S_{x,z}$ and the polarization component
$e_{x}$ invariant. Another restriction is the absence of total magnetic moment of the sample which means
that both $S_x$ and $S_z$ must satisfy to the requirement of zero net polarization:

\begin{equation}
\int_{0}^a S_{x}(x) dx = \int_{0}^a S_{z}(x) dx = 0.
\label{net}
\end{equation}

The spin density distribution in the superlattice cell is found for three different
polarizations: linear along $x$ axis, linear along $y$ axis, and circular in the $(xy)$ plane
of 2DEG. The Fermi level in Fig.\ref{figband} varies from $4$ to $20$ meV counted from the bottom of
the topmost partially filled electron size quantization band. According to the spectrum
in Fig.\ref{figband}, this variation fills gradually all four minibands shown there.
The excitation energy is chosen to be $\hbar \omega=10$ meV which corresponds
to $\omega/2\pi=2.43$ THz and provides effective transitions between the occupied and vacant minibands,
as it can be seen in Fig.\ref{figband}.

\begin{figure}[ht]
  \centering
  \includegraphics[width=80mm]{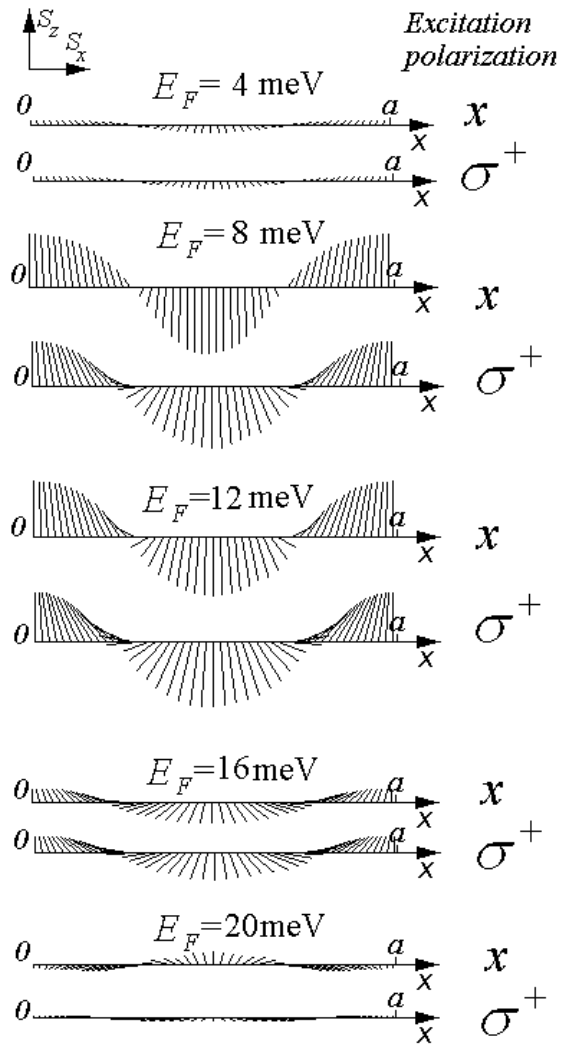}
  \caption{Excited spin density distribution  $(S_x(x),S_z(x))$ along the 1D superlattice elementary cell
           created under $x$- and $\sigma^{+}$-polarized terahertz excitation with
           the frequency $\omega/2\pi=2.43$ THz and with the intensity $I=0.3$ $W/\rm{cm}^{2}$.}
  \label{figxp}
\end{figure}

\begin{figure}[ht]
  \centering
  \includegraphics[width=80mm]{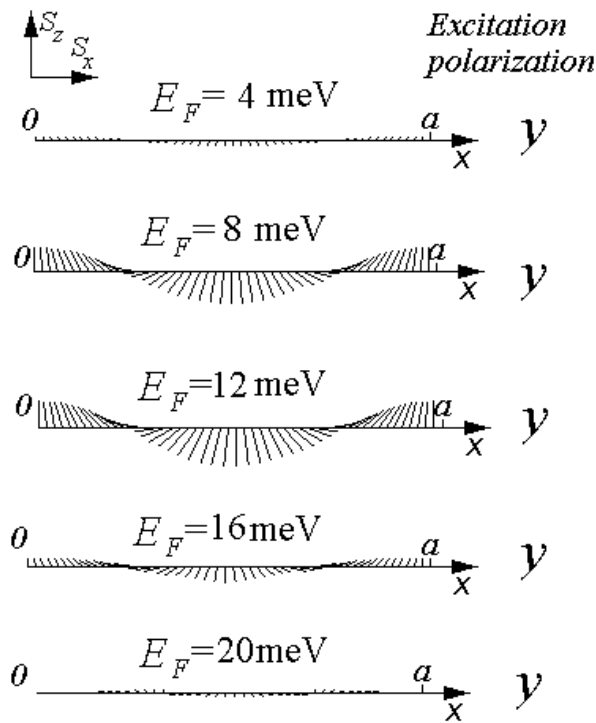}
  \caption{Excited spin density distribution  $(S_x(x),S_z(x))$ along the 1D superlattice elementary cell
           created under $y$-polarized terahertz excitation with the same parameters as in Fig.\ref{figxp}
           but at higher intensity $I=0.9$ $W/\rm{cm}^{2}$.}
  \label{figy}
\end{figure}

First, let us consider the cases of the excitation linearly polarized along $x$ axis and
the $\sigma^{+}$- circular polarized one. The excited non-equilibrium spin density (\ref{s}) is shown
in Fig.\ref{figxp} as a 2D vector field  $(S_x(x),S_z(x))$. Its magnitude is proportional to the radiation
intensity which is $0.3$ $W/cm^2$ for the present case and the texture shape is varied
with respect to the Fermi energy. The magnitude of the arrow length in Fig.\ref{figxp}, i.e.
the maximum excited spin density can be obtained from the value of the excited charge
density $n_{\rm{ex}}$. Taking the data from the experiments with optical excitation\cite{Olesberg}
where the volume concentration of the excited carriers reached $10^{16} \rm{cm}^{-3}$
one can estimate the excited surface concentration $n_{\rm{ex}}$ to be of the order of
$10^{10} \ldots 10^{11}$ $\rm{cm}^{-2}$.
One can see in Fig.\ref{figxp} that the spin textures are similar for $x$- and $\sigma^{+}$
radiation, as well as for $\sigma^{-}$ (not shown). The explanation is that all these polarizations contain
the $x$ component of transition matrix elements which causes the most effective transitions in
the $x$-oriented superlattice. The transformation of spin density distribution with the Fermi level
position in Fig.\ref{figxp} is produced by gradual filling of the minibands. The small amplitudes of
spin density (at $E_F=4$ meV in Fig.\ref{figxp}) correspond to the small filling factor
(see Fig.\ref{figband}). By increasing the Fermi energy the complete filling of two lowest subbands
(at $E_F=8$ meV and $E_F=12$ meV in Fig.\ref{figxp}) is reached and, finally, the excited spin density
magnitude decreases again with complete filling of all four nearest miniband in Fig.\ref{figband}
(at $E_F=16$ meV and $E_F=20$ meV in Fig.\ref{figxp}).
It can be seen in Fig.\ref{figxp} that the excited spin density texture have the primary spatial wavelength
being close to the superlattice period $a$. This feature can be explained by the structure
of the matrix element of the transitions  which reaches maximum amplitude at $k_x=\pm \pi/a$,
leading to the effective creation of spin texture (\ref{s}) with the specific primary wavelength
equal to $a$. Since the Fermi level position can be varied in experiments by tuning the gate voltage
which controls the concentration of 2DEG, our model predicts a possible and realistic mechanism for
optical creation of various spin textures.

Another type of the considered polarization is the radiation polarized
along $y$ axis. The structure is homogeneous in this direction and the only reason for
the transition probabilities to be nonzero is the non-parabolic character of energy
spectrum as a function of $k_y$ due to the interplay between SO and superlattice
potential. However, this interplay is reduced with increasing $k_y$ and thus one can
expect the probabilities to be smaller for the $y$-polarized radiation than
for the $x$-polarized one, making the creation of the excited spin texture comparable to the one in
Fig.\ref{figxp} possible at higher intensities. This suggestion is confirmed by the spin textures
in Fig.\ref{figy} where the similar spin textures as in Fig.\ref{figxp} are created at
the intensity $0.9$ $W/\rm{cm}^2$ which is three times greater than for the $x$- and $\sigma$- polarized light.
Nevertheless, all of the intensities considered in the paper are within the range of
$0.5 - 1$ $W/\rm{cm}^2$ which is accessible in modern experimental
setups.\cite{Ganichev07,Yang,Cho,Zhao,Olesberg}

Further investigations of gated 2DEG with SO interaction would require the studies of
the spin current conventionally described by the operator
${\hat j}^s_{ij}=\hbar \{ {\hat v_i},{\sigma_j} \}/4$ where
$\hat v_i=\partial {\hat H}/ \partial k_i$. The excited spin current distribution can
be analyzed under the same approach as the spin density, and the spin separation distance
can be calculated.\cite{Bhat} The investigations of spin current in our system deserves a separate
detailed discussion and will be performed in the forthcoming paper.

\section{Conclusions}

We have studied the excited spin texture distribution in 2DEG with Rashba spin-orbit interaction
subject to 1D tunable superlattice potential and illuminated by the terahertz radiation with
different polarizations and intensities. It was found that in the absence of the net polarization
the local excited spin texture can be effectively manipulated by varying the Fermi level position
in 2DEG as well as the intensity and polarization of the radiation at fixed terahertz frequency.
The effect of excited spin texture creation discussed in the paper has a qualitative
character and should be observable in a wide class of two-dimensional heterostructures
where the spin-orbit coupling energy is more pronounced than the temperature broadening or
the smearing caused by edges, defects or impurities.

\section*{Acknowledgments}

The author thanks V.Ya. Demikhovskii and A.A. Perov for many helpful discussions.
The work was supported by the RNP Program of the Ministry of Education and Science RF,
by the RFBR, CRDF, and by the Foundation "Dynasty" - ICFPM.

\end{document}